\documentclass[prb,twocolumn,superscriptaddress]{revtex4}
\usepackage{epsfig}

\begin{document}
\title{Magneto-Transport Properties of Doped RuSr$_2$GdCu$_2$O$_8$}

\author{J.E. McCrone}\address{IRC in Superconductivity, Cambridge University, Cambridge
CB3 0HE, U.K.}\author{J.L. Tallon} \address{Industrial Research
Ltd., P.O. Box 31310, Lower Hutt, New Zealand.}\author{J.R.
Cooper}\address{IRC in Superconductivity, Cambridge University,
Cambridge CB3 0HE, U.K.}
\author{A.C. MacLaughlin}\address{IRC in Superconductivity and
Department of Chemistry, Cambridge University, Cambridge CB2 1EW,
U.K.}\author{J.P. Attfield}\address{IRC in Superconductivity and
Department of Chemistry, Cambridge University, Cambridge CB2 1EW,
U.K.}\author{C. Bernhard} \address{Max-Planck-Institut f\"ur
Festk\"orperforschung, D-70569 Stuttgart, Germany.}
\date{\today}

\begin{abstract}
RuSr$_2$GdCu$_2$O$_8$, in which magnetic order and
superconductivity coexist with $T_{Magnetic}$$\gg$$T_c$, is a
complex material which poses new and important questions to our
understanding of the interplay between magnetic and
superconducting (SC) order. Resistivity, Hall effect and
thermopower measurements on sintered ceramic RuSr$_2$GdCu$_2$O$_8$
are presented, together with results on a broad range of
substituted analogues. The Hall effect and thermopower both show
anomalous decreases below $T_{Magnetic}$ which may be explained
within a simple two-band model by a transition from localized to
more itinerant behavior in the RuO$_2$ layer at $T_{Magnetic}$.
\end{abstract}


\maketitle

\section*{Introduction}
Soon after the first successful synthesis of RuSr$_2$GdCu$_2$O$_8$
\cite{Bauernfeind1996} the material was found to display not only
superconductivity ($T_c\simeq$45K) but coexisting magnetic order
with $T_{Curie}\simeq$135K\cite{Tallon1999,Bernhard1999}.
Evidence accumulated from static magnetization, muon spin rotation
($\mu$SR)\cite{Bernhard1999} and from Gd-ESR\cite{Fainstein1999}
studies showing that the magnetism is a spatially uniform bulk
property. Specific heat measurements\cite{Tallon1999} and the
diamagnetic shielding fraction at low
temperatures\cite{Bernhard1999,Bernhard2000,McCrone2000_2}
indicate that the superconductivity is also a bulk property, and
that the two phases therefore coexist on a truly microscopic
scale. An initial neutron diffraction study eliminated the
possibility of ferromagnetic order with the Ru moments lying in
the RuO$_2$ plane, but did not rule out ferromagnetic alignment
with the moments parallel to the $c$-axis, canted ferromagnetism
or itinerant ferromagnetism\cite{Chmaissem2000}.  Subsequent
polarized neutron diffraction data\cite{Lynn2000} have thrown the
debate on RuSr$_2$GdCu$_2$O$_8$ wide open by appearing to show
that the underlying ordering of the Ru moments below the magnetic
transition is in fact G-type antiferromagnetic (antiparallel
nearest-neighbour ordering in all three crystallographic
directions). Finally, the most recent neutron measurements on
RuSr$_2$YCu$_2$O$_8$ confirmed that there is indeed a
ferromagnetic component of about 0.28 $\mu_B$ which is about
1/5$^{\normalfont{th}}$ of the AF component of 1.2 $\mu_B$
\cite{Takigawa}. The magnetic order shows a rather strong and
unusual response to an applied magnetic field, with the FM
component growing rapidly in strength and dominating over the AF
already at 2T. Whatever the nature of its magnetism, the discovery
of this material is an exciting development which poses new and
important questions to our understanding of the interplay between
magnetic and SC order.

Magnetoresistance (MR), Hall effect and thermopower (TEP)
measurements on undoped sintered ceramic RuSr$_2$GdCu$_2$O$_8$
were presented previously\cite{McCrone1999}.  Above $T_{Magnetic}$
the MR is negative and proportional to the square of the Ru
magnetization and was ascribed to spin-scattering of the current
carriers. A model for dilute magnetic alloys was used to extract
from these data a value ($\geq$25 meV) for the exchange
interaction between the Ru moments and the carriers.  Below
$T_{Magnetic}$ the Hall effect and TEP both fall anomalously. It
will be shown that these data may be explained within a simple
two-band model by a transition from localized to more itinerant
behavior in the RuO$_2$ layer at $T_{Magnetic}$. Evidence for
delocalized carriers within the RuO layers has also been obtained
from other transport and microwave absorption
studies\cite{Williams2002} as well as from Ru-NMR measurements
where clear anomalies in the Ru-NMR relaxation rate occur near
T$_c$ \cite{Tokunaga}. This observation suggests that the Ru
nuclear moments experience a sizeable hyperfine coupling to the
charge carriers that enter the SC state.

The magneto-thermopower reveals an extremely unusual variation of
$T_c$ with applied field\cite{McCrone2000_2}: $T_c$ actually
\textit{increases} by $\sim 4K$ as the applied field is increased
to 2T.  The increase saturates along with the Ru magnetization,
suggesting that the onset of Ru magnetic order reduces a magnetic
pair-breaking effect in the CuO$_2$ layer.

The carrier concentration in RuSr$_2$GdCu$_2$O$_8$ and its
magnetic and SC properties, structural deformations, and so forth
may be altered by cation substitution. Examining the transport
properties of such samples should lead to a better understanding
of the parent material.  In this paper we present
magneto-transport measurements on substituted
RuSr$_2$GdCu$_2$O$_8$.  It will be shown that the data strongly
support a simple two-band model in which the Hall effect and TEP
of each sample are determined by the properties of the CuO$_2$ and
RuO$_2$ layers, weighted appropriately by their conductivities.
The model indicates that the RuO$_2$ layer in the undoped material
is very poorly conducting at room temperature, with
$\sigma_{Ru}\sim 0.1\sigma_{Cu}$, increasing to $\sim
0.3\sigma_{Cu}$ or higher at low temperature.  While in most of
the samples studied the CuO$_2$ layer remains the better conductor
at all temperatures, we find that the RuO$_2$ layer dominates the
conductivity below $T_{Magnetic}$ in a sample with 10\% Ce$^{4+}$
substituted for Gd$^{3+}$.

\section*{Experimental Methods}
Phase-pure sintered pellets of RuSr$_2$GdCu$_2$O$_8$ were
synthesized as described previously via solid-state reaction of a
stoichiometric mixture of high-purity metal oxides and
SrCO$_3$\cite{Tallon1999_2,Bernhard1999}.  The doped samples,
listed in Table I, were produced similarly; the compositions given
are nominal.  A final extended anneal at 1060$^\circ$C in flowing
high-purity O$_2$ produces a marked improvement in the
crystallinity of the undoped material, resulting in a higher
resistive $T_c$ (as defined by $\rho(T)=0$) but no significant
change in the thermodynamic T$_c$\cite{Tallon1999}.
\begin{table}
\narrowtext
\begin{tabular}{@{\hspace{0.1in}}ll@{\hspace{0.1in}}}
\textbf{Composition} & \textbf{Substituted site}\\\hline
Ru$_{0.6}$Sn$_{0.4}$Sr$_2$GdCu$_2$O$_8$& 40\% Sn for Ru\\
Ru$_{0.8}$Sn$_{0.2}$Sr$_2$GdCu$_2$O$_8$& 20\% Sn for Ru\\
Ru$_{0.925}$Sn$_{0.075}$Sr$_2$GdCu$_2$O$_8$& 7.5\% Sn for Ru\\
Ru$_{0.975}$Sn$_{0.025}$Sr$_2$GdCu$_2$O$_8$& 2.5\% Sn for Ru\\
Ru$_{0.8}$Nb$_{0.2}$Sr$_2$GdCu$_2$O$_8$& 20\% Nb for Ru\\
Ru$_{0.9}$Nb$_{0.1}$Sr$_2$GdCu$_2$O$_8$& 10\% Nb for Ru\\
RuSr$_2$Gd$_{0.8}$Ce$_{0.2}$Cu$_2$O$_8$& 20\% Ce for Gd\\
RuSr$_2$Gd$_{0.9}$Ce$_{0.1}$Cu$_2$O$_8$& 10\% Ce for Gd\\
RuSr$_2$EuCu$_2$O$_8$& 100\% Eu for Gd\\
RuSr$_2$Gd$_{0.6}$Dy$_{0.4}$Cu$_2$O$_8$& 40\% Dy for Gd\\
RuSr$_2$Gd$_{0.9}$Y$_{0.1}$Cu$_2$O$_8$& 10\% Y for Gd\\
RuSr$_2$GdCu$_{1.9}$Li$_{0.1}$O$_8$& 5\% Li for Cu
\label{table:doped}
\end{tabular}
\caption{Substituted variants of RuSr$_2$GdCu$_2$O$_8$ studied in
this work.} \widetext
\end{table}

Bars of approximate dimensions 4$\times$1$\times$0.7mm$^3$ were
cut from the sintered pellets using a diamond wheel, then polished
down to a thickness of $\sim 150\mu$m in order to increase the
measured Hall voltage.  They were mounted on quartz substrates in
a standard six-contact configuration allowing both resistance and
Hall voltage to be measured simultaneously.  The contacts were
made using 25$\mu$m gold wire and \textit{Dupont 6838} conducting
epoxy, cured in air at 450$^\circ$C for six minutes, giving
contact resistances $<1\Omega$.

Resistivity and Hall effect measurements were made using an ac
current source, low-noise transformers and lock-in amplifiers. A
frequency of $\sim77$Hz was used to avoid mains pick-up, with
current densities of around 0.25Acm$^{-2}$.  The Hall coefficient,
$R_H$, was usually measured by stabilizing the temperature and
field (10T unless stated otherwise), then measuring the Hall
voltage with the sample rotated by 0$^\circ$\ and 180$^\circ$\
with respect to the field. The Hall coefficient is then given by
$R_H={{\left(V_0-V_{180}\right)t}\over{2\,IB}}$ where \textit{B}
is the magnetic field, \textit{t} the sample thickness and
\textit{I} the current.  This method eliminates the MR of the
sample, and the offset voltage from $\rho_{xx}$ due to contact
misalignment.  Where $R_H$ was measured as a function of field,
this was swept to both positive and negative values and $R_H(B)$
determined from $V_{B}-V_{-B}$.

TEP measurements were made by the `toggled' heating
method\cite{Cooper1971,Resel1996}. Two 25$\mu$m chromel-alumel
thermocouples, attached to the sample with small blobs of silver
paint, measure both the thermal emf and temperature gradient,
ensuring that these are measured between the same two points.  The
sample is first stabilized at the measurement temperature, a small
thermal gradient is applied, and the resulting thermal emf
measured.  The thermal gradient is then reversed, allowing slowly
changing thermal emfs in the cryostat wires to be nulled out. A
`rest state' was added whereby both ends of the sample were heated
at half-power, providing two extra measurement points. Adding this
state keeps the total power dissipation into the stage constant,
avoiding fluctuation of its temperature when the heater currents
are changed.


\section*{Results}
\subsection*{Transport Measurements on Pure RuSr$_2$GdCu$_2$O$_8$}
Hall effect, thermopower and resistivity data for undoped
RuSr$_2$GdCu$_2$O$_8$ are shown in Fig. \ref{fig:figure1}.  The
room-temperature value of the TEP implies a hole concentration,
$p_{Cu}$, of 0.06-0.07 holes/Cu\cite{Obertelli1992}, while its
temperature dependence is typical of other high-$T_c$ materials,
with the exception of the unusual linear temperature dependence
below $T_{Magnetic}$. The overall magnitude and temperature
dependence of the Hall coefficient is consistent with a doping
level, $p_{Cu}$, of $\simeq 0.07$ holes/Cu, as inferred from the
room-temperature TEP. $R_H$ displays a high-$T_c$-like temperature
dependence well above $T_{Magnetic}$. However, below about 170K
there is an anomalous downturn in $R_H$ which is not seen in
typical high-$T_c$ data. The so-called `anomalous' Hall effect
observed in magnetic materials has been measured and discounted as
the cause of this downturn\cite{McCronethesis}. Alternative
possibilities are that it is due to charge delocalization in the
RuO$_2$ plane occurring near the magnetic transition, or to charge
transfer into the CuO$_2$ layers. It will be shown that a two-band
model, with a localized to itinerant transition occurring at
$T_{Magnetic}$ in the RuO$_2$ layer, can explain both these and
the TEP data.

\subsubsection*{The conductivity of the RuO$_2$ layer}
We now introduce a simple two-band model which successfully
describes the features of the above data.  The bands in this model
are those formed by carriers in the Cu and Ru orbitals; the
overall TEP and Hall effect are given by the sum of the CuO$_2$
and RuO$_2$ layer values, weighted by the layer
conductivities\cite{Hurd1972,Barnard}:
\begin{equation}
R_H={{R_H^{Ru}\,\left({\sigma^{Ru}_{xx}}\right)^2 +
R_H^{Cu}\,\left({\sigma^{Cu}_{xx}}\right)^2} \over
\left({\sigma^{Ru}_{xx}+\sigma^{Cu}_{xx}}\right)^2}
\label{eqn:h_add}
\end{equation}
\begin{equation}
S={{S^{Ru}\,\sigma^{Ru}_{xx} + S^{Cu}\,\sigma^{Cu}_{xx}} \over
{\sigma^{Ru}_{xx}+\sigma^{Cu}_{xx}}} \label{eqn:s_add}
\end{equation}
With some reasonable estimates of the RuO$_2$ and CuO$_2$ layer
properties, it is possible to use this model and the measured
room-temperature Hall effect and TEP to put a limit on the
conductivity of the RuO$_2$ layer.  To do this we assume that the
Hall coefficient of the RuO$_2$ layer is approximately zero (the
maximum value observed in other two-dimensional Ru oxides studied
to date is $5\times
10^{-10}$m$^3$C$^{-1}$)\cite{Yang1999,Mackenzie1996,Perry2000}.
With this assumption in Eqn. \ref{eqn:h_add}, the conductivity of
the RuO$_2$ layer may be estimated from
\begin{equation}\begin{array}{l}
{\sigma^{Ru}_{xx}\over \sigma^{Cu}_{xx}} = \sqrt{{R_H^{Cu}\over
R_H}}-1\\ \\ R_H^{Ru}\ll R_H^{Cu};\,\,
\sigma^{Ru}_{xx}<\sigma^{Cu}_{xx} \label{eqn:halladd3}
\end{array}\end{equation}
The ratio of the Hall coefficient of the CuO$_2$ layers $R_H^{Cu}$
to the measured value $R_H$ caused by the presence of the RuO$_2$
layer, is hard to estimate due to the uncertain doping state in
RuSr$_2$GdCu$_2$O$_8$ and the spread of values of $R_H^{Cu}$, for
a given doping level, in the
literature\cite{Wuyts1996,Carrington1993,Fisher1994}. Given these
uncertainties, a reasonable range of values of $R_H^{Cu}/R_H$ is
1-1.4, giving $\sigma^{Ru}_{xx}$ in the range
$0-0.18\,\sigma^{Cu}_{xx}$. The summary of $R_H^{Cu}$ values in
the review by Cooper and Loram\cite{Cooper1996} would favor the
low end of this range.

For this range of conductivity in the RuO$_2$ layer, Equation
\ref{eqn:s_add} predicts that the measured net TEP lies some
0-8$\mu$VK$^{-1}$ below the intrinsic CuO$_2$ layer value, i.e.
that $75\le S_{290}^{Cu}\le 83$. It is very unlikely that
$S_{290}^{Cu}$ lies in the upper half of this range: a value of
$S=83\mu$VK$^{-1}$ would imply an extremely small hole
concentration for which a $T_c$ as high as 46K would be
extraordinary.

Having placed a limit on the conductivity one can use a 2D model
to determine $k_Fl$, the product of the Fermi wave-vector with the
mean free path for the RuO$_2$ layers. This quantity gives an
indication as to whether the carriers are localized or itinerant
and for a cylindrical Fermi surface may be written as

\begin{equation}
k_Fl = \sigma{2\pi\hbar c \over e^2}
\end{equation}
where $c$ is the separation of the planes. Data in the literature
for the \textit{ab}-plane resistivity of under-doped
YBa$_2$Cu$_3$O$_{7-\delta}$ (YBCO) films and single crystals, with
$p\simeq 0.07$, give a consistent value of 1.2m$\Omega$cm at room
temperature \cite{Carrington1993,Sun1994,Jin1998,Carrington1992}
giving $k_Fl_{Cu}=1.3$, near the limit of localization. In fact,
in only slightly more under-doped samples one sees a
semiconducting upturn at low temperatures. Given the range of
ratios of $\sigma^{Ru}_{xx}$ to $\sigma^{Cu}_{xx}$ derived from
the Hall effect, $k_Fl_{Ru}=0-0.45$ at room temperature. The TEP
data suggest that the true value is at the low end of this range,
indicating that the carriers in the RuO$_2$ layers are at best
very poorly metallic.

\subsubsection*{Temperature dependence of $\sigma_{Ru}$}
Having established that the room-temperature conductivity of the
RuO$_2$ layer is close to zero, typical $S(T)$ and $R_H(T)$ data
for high-$T_c$ superconductors may be scaled so that the room
temperature values match those of RuSr$_2$GdCu$_2$O$_8$.  The
differences below $T_{Magnetic}$ may then be used to follow
$\sigma_{Ru}$ as a function of temperature.

\begin{figure}[htb]
\begin{center}
  \epsfig{file=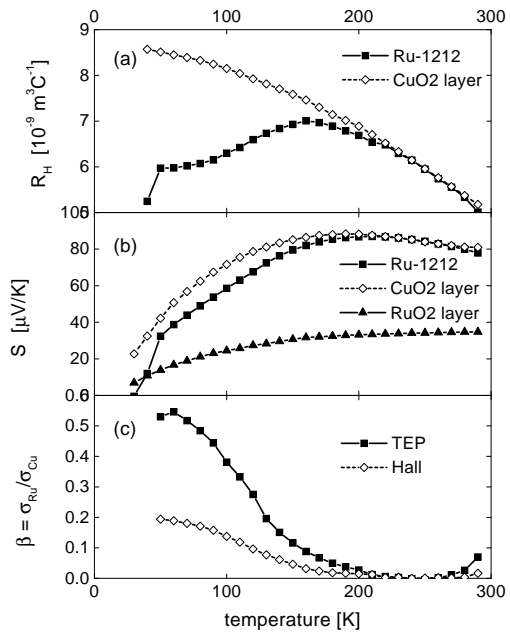, width=2.8in}
  \caption{(a) Hall effect and (b) thermopower data for RuSr$_2$GdCu$_2$O$_8$, together with
  estimated values of the CuO$_2$ and RuO$_2$ layer properties as described in
  the text.  Panel (c) shows the ratio
  $\beta={\sigma_{Ru}\over\sigma_{Cu}}$ calculated in the two-band model,
  assuming $\sigma_{Ru}\simeq 0$ well above $T_{Magnetic}$.}
  \label{fig:figure1}
\end{center}
\end{figure}

Typical Hall effect data for the CuO$_2$ layer have been taken
from measurements on sintered Ca-doped
YBa$_2$Cu$_3$O$_{7-\delta}$, while $R_H^{Ru}$ will be set to zero,
its value in other RuO$_2$ layer compounds being much lower than
$R_H^{Cu}$\cite{Yang1999,Mackenzie1996,Perry2000}. Typical
$S^{Cu}$ data are approximated by measurements on sintered
YBa$_2$Cu$_3$O$_{7-\delta}$ with $\delta = 0.53$\cite{Cooper1996},
multiplied by 1.12 to match the high-temperature
RuSr$_2$GdCu$_2$O$_8$ data.  Finally, $S^{Ru}$ is approximated by
data measured on a sintered sample of SrRuO$_3$, which displays a
magnitude and temperature dependence similar to that of
Sr$_2$RuO$_4$\cite{Yoshino1996}. All these data are shown in
Figure \ref{fig:figure1}, together with the resulting
$\beta(T)={\sigma_{Ru}\over\sigma_{Cu}}$ calculated from Eqns.
\ref{eqn:h_add} and \ref{eqn:s_add}.

Given the uncertainties in the approximated RuO$_2$ and CuO$_2$
layer properties the two $\beta(T)$ curves calculated
independently from the drops in $S(T)$ and $R_H(T)$ agree well
qualitatively.  If TEP data for a sample of 20\% Sn-doped
RuSr$_2$GdCu$_2$O$_8$, in which we shall argue that $\sigma_{Ru}$
is strongly suppressed below $T_{Magnetic}$, is used to
approximate $S^{Cu}$, the agreement is also quantitative.  Because
the TEP is a less sensitive function of $\beta$ than the Hall
effect, the difference between $S^{1212}$ and $S^{Cu}$ is quite
small compared with that between $R_H^{1212}$ and $R_H^{Cu}$. Thus
the value of $\beta$ calculated from the TEP data is more
sensitive to inaccuracy in the assumed $S^{Cu}$ data.  This
explains why using the (only slightly different) 20\% Sn-doped
data to approximate $S^{Cu}(T)$ results in a better match to
$\beta(T)$ calculated from the Hall effect. Whichever data are
used, the results show a rapid rise in the relative conductivity
of the RuO$_2$ layer below 150K, to $\sim 0.3\sigma_{Cu}$ or
higher.

\subsection*{Transport Measurements on Substituted RuSr$_2$GdCu$_2$O$_8$}
\subsubsection*{Sn-Doped RuSr$_2$GdCu$_2$O$_8$}
The diamagnetic Sn$^{4+}$ ion substitutes for Ru in solid
solution, and is slightly larger in size than Ru$^{4+/5+}$.  The
effects of doping the Ru site are of extreme interest given the
current debate regarding the spin and charge configuration of the
Ru ions\cite{Williams2000,Butera2000,Liu2000}.

We note that the Sn-doped samples studied here were from two
sources prepared with slightly different annealing strategies.
Comparison of their sample resistivities is therefore not
necessarily meaningful, as annealing strongly affects the
grain-boundary conductivity of
RuSr$_2$GdCu$_2$O$_8$\cite{McCrone1999}.  In general the
resistivity of sintered high-$T_c$ materials is also affected by
sample density\cite{Cooper1991,Fisher1996}.

\begin{figure}[htb]
\begin{center}
  \epsfig{file=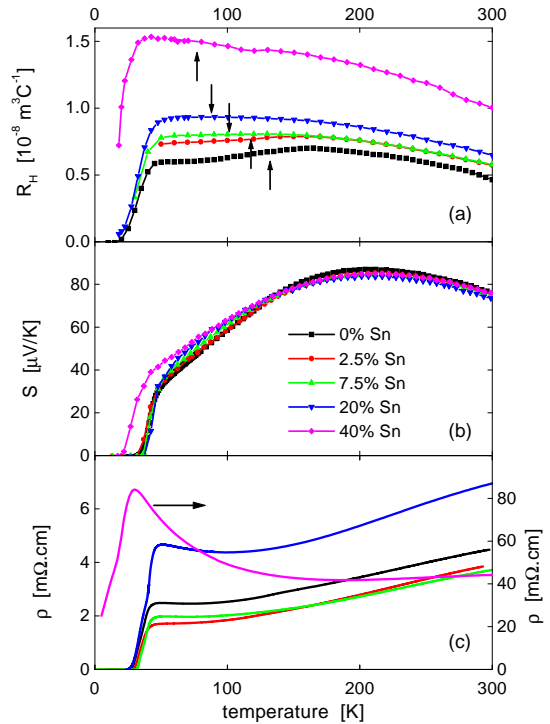, width=3.2in}
  \caption{(a) Hall effect, (b) thermopower and (c) resistivity data for
  Sn-doped samples of RuSr$_2$GdCu$_2$O$_8$.}
  \label{fig:figure2}
\end{center}
\end{figure}

For the 2.5\% and 7.5\% samples the resistivity (Figure
\ref{fig:figure2}) is metallic, and similar in magnitude to the
undoped sample.  The 20\% sample has a higher resistivity and
shows a small semiconducting upturn at low temperatures, while
both the magnitude and upturn are far larger for the 40\% sample.
Estimating $T_c$ from the onset of the resistive transition
reveals a gradual increase from 40.5K for the 2.5\% sample to
43.5K for the 20\% sample, while the 40\% sample has a reduced
$T_c$ of just 30K.

The TEP, $S(T)$, and Hall effect, $R_H(T)$, are much less affected
by grain-boundaries than the resistivity. In conventional
high-T$_c$ materials they closely reflect bulk CuO$_2$ layer
properties in conventional high-$T_c$
materials\cite{Carrington1994,Fisher1996}.
The Hall effect data show a slow and monotonic decrease in $T_c$
with increasing Sn concentration, but it should be remembered that
these data were taken in a field of 10T and only partly reflect
the zero-field $T_c$. The vertical arrows in Figure
\ref{fig:figure2} (and in subsequent figures) show the location of
the magnetic transition. The TEP data show that $T_c$ (defined by
the maximum in the derivative) rises by $\sim$4K in going from the
2.5\% sample to the 20\% sample, in good agreement with the
resistivity data. The 40\% sample shows a much lower transition
temperature, both in $R_H$ and $S$.  The increase in $T_c$ with Sn
concentration is attributed to a transfer of holes into the
CuO$_2$ layer\cite{Maclaughlin1999_2}, though we observe a smaller
increase than the $\sim$12K reported
previously\cite{Maclaughlin1999_2,Maclaughlin2000}.  In the
earlier studies $T_c$ was defined from the resistivity onset, and
the $T_c$ values obtained for low doping levels were significantly
lower, possibly due to granularity.

On examining the temperature and doping dependence of the
normal-state properties, one immediately observes that the room
temperature TEP $S_{290}$ is little changed by the addition of Sn.
This result is strange given the rise of $\sim4$K in $T_c$ as the
doping level is increased to 20\%. The change in the Hall effect
is also counter-intuitive: the 30\% increase in going from 0 to
20\% Sn would normally indicate a \textit{decrease} in hole
concentration. This apparent paradox is resolved when it is
noticed that the anomalous drop in $R_H$ below $T_{Magnetic}$ is
diminished in the 2.5\% and 7.5\% samples, and is absent in the
20\% sample: as the Sn concentration is increased the RuO$_2$
layer becomes well localized below $T_{Magnetic}$, reflecting
significantly reduced conductivity at all temperatures. The
changes in $R_H(290)$ and $S_{290}$ may then be explained quite
simply: the introduction of Sn dopes a few extra holes into the
CuO$_2$ layer, increasing $p_{Cu}$ and raising $T_c$ by $\sim$4K,
but also drives the RuO$_2$ layer more insulating. Thus while
$R_H^{Cu}$ probably decreases slightly, the overall Hall effect
increases as the RuO$_2$ layer no longer provides a parallel
conduction pathway. The slight increase in $p_{Cu}$ which would
normally decrease the measured TEP, is balanced by the decreasing
$\sigma_{Ru}$, which removes the suppression of the TEP by the
RuO$_2$ layer, leaving it relatively unchanged overall. Certainly,
the increase in doping is far smaller than one would expect from
substituting Sn$^{4+}$ for Ru$^{5+}$, suggesting that the mean
valency of the Ru ion is less than 5+.  This conclusion is
supported by recent XANES measurements\cite{Liu2000}.

The reduction in the room-temperature $R_H$ of pure
RuSr$_2$GdCu$_2$O$_8$, due to conductivity in the RuO$_2$ layer,
was estimated to be of the order of 30\%.  This is entirely
consistent with the rise in $R_H$ observed as the Sn concentration
is increased to 20\%, assuming that $\sigma_{Ru}\rightarrow 0$.
The 40\% Sn-doped sample does not fit well into this picture,
having a much larger $R_H$ at all temperatures. Given the much
larger resistivity of this sample and its drastically reduced
$T_c$, it is possible that some Sn$\leftrightarrow$Cu substitution
has occurred, reducing the CuO$_2$ layer doping state, or that
there are significant impurities present.

\subsubsection*{Nb-Doped RuSr$_2$GdCu$_2$O$_8$} 
\begin{figure}[htb]
\begin{center}
  \epsfig{file=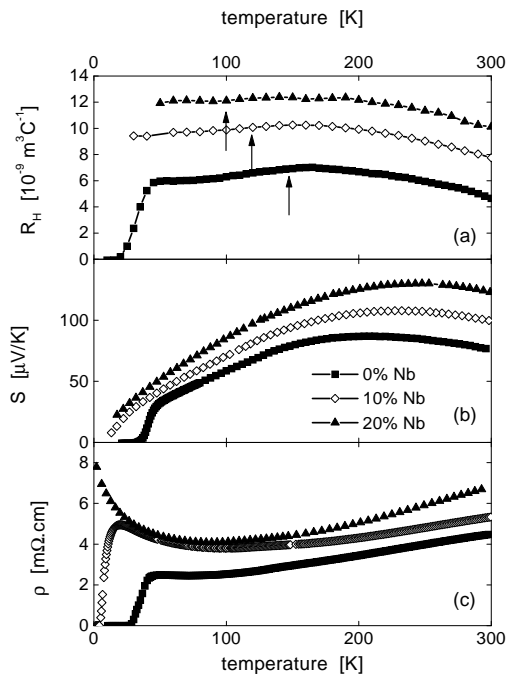, width=3.2in} \caption{(a) Hall effect,
  $R_H(T)$, (b) thermopower, $S(T)$, and (c) resistivity,
  $\rho(T)$, data for Nb-doped RuSr$_2$GdCu$_2$O$_8$.}
  \label{fig:figure3}
\end{center}
\end{figure}

Nb also substitutes for Ru in the
RuSr$_2$GdCu$_2$O$_8$ structure, but has a dramatically different
effect on the transport properties.  In contrast to the Sn ion,
which has a charge of 4+, Nb is believed to substitute in its
usual 5+ state\cite{Maclaughlin2000}, and thus for an average Ru
valency of less than five will remove holes from the system,
further under-doping it. The room-temperature TEP bears this out,
showing a large increase proportionate with Nb doping (see Figure
\ref{fig:figure3}) and confirming that the CuO$_2$ layer is
progressively under-doped by the substitution of Nb. This
conclusion is supported by the commensurate increase in the Hall
effect and the rapid reduction of $T_c$, which is 19K for the 10\%
sample and below 1.5K (if present at all) in the 20\% sample.

The effect of Nb doping on the Ru layer is less clear.  The Hall
effect of the 10\% sample displays a maximum near $T_{Magnetic}$,
suggesting increased RuO$_2$ conductivity below this temperature,
but the drop is not as clear as in the undoped sample.  For 20\%
Nb$\leftrightarrow$Ru substitution $R_H$ rises to a more or less
constant value of 1.2$\times 10^{-8}$m$^3$C$^{-1}$ below 200K, and
there is no sign of a significant change at $T_{Magnetic}$. The
conclusion from the TEP and Hall effect data, then, is that the
transition from localized to itinerant behavior of the RuO$_2$
layer is suppressed by the addition of Nb, as it is by the
addition of Sn.

The resistivity, on the other hand, shows surprisingly little
difference between the 10\% and 20\% samples -- in fact the
residual resistivity (extrapolated from the linear
high-temperature data) actually \textit{decreases}.  A possible
scenario consistent with this result is that the RuO$_2$ layer
becomes \textit{more} itinerant both above and below
$T_{Magnetic}$ as the Nb level is increased.  However, if this
were the case, the increased $\sigma_{Ru}$ would be expected to
suppress both $R_H$ and $S$ below the CuO$_2$ plane values.  In
fact, for $T_c$=19K and $T_c^{max}\approx 100$K, the universal
relationship between $S_{290}$ and $T_c$ predicts
$S_{290}^{Cu}\sim 100\mu$VK$^{-1}$, as observed.  Thus, while the
increased $S$ and $R_H$, and the reduced $T_c$ are consistent with
a reduced hole concentration in the CuO$_2$ layer and a localized
RuO$_2$ layer, the relatively good conductivity of the 20\%
Nb-doped sample is not. One possible explanation is that the
behavior of the resistivity is extrinsic to the bulk in the 20\%
Nb sample, resulting from either increased grain-boundary
conductivity, or increased sample density.

\subsubsection*{Ce-Doped RuSr$_2$GdCu$_2$O$_8$} Unlike Nb and Sn, which substitute
for Ru, Ce substitutes for Gd in the layer separating the two
CuO$_2$ planes, and so would be expected to affect these more than
the RuO$_2$ layers from which it is relatively remote. The Ce ion
is expected to be in the 4+ state in RuSr$_2$GdCu$_2$O$_8$, as it
is in the structurally similar compound
RuSr$_2$(Gd$_{1+x}$Ce$_{1-x}$)Cu$_2$O$_{10}$\cite{Felner1999};
hence its substitution for Gd$^{3+}$ should further underdope the
material.

\begin{figure}[htb]
\begin{center}
  \epsfig{file=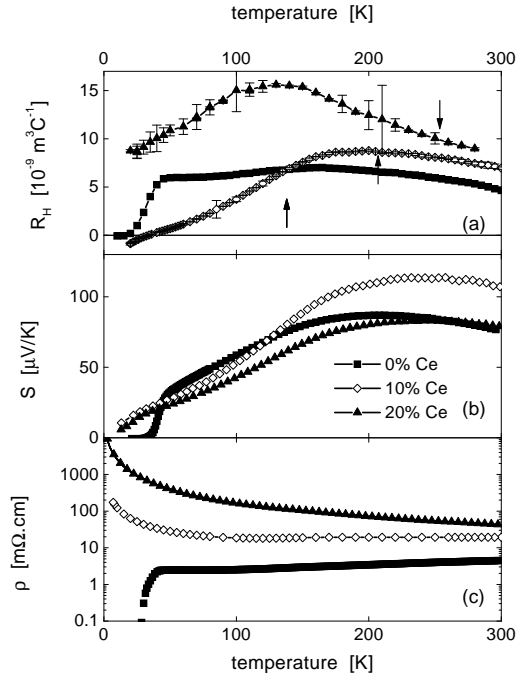, width=3.2in}\caption{(a) Hall effect,
  $R_H(T)$, thermopower, $S(T)$, and Resistivity, $\rho(T)$, data
  for Ce-doped RuSr$_2$GdCu$_2$O$_8$.  The large error bars shown
  on the 20\% Ce Hall data points result from the exceptionally
  large resistivity of the sample making balancing difficult at
  low temperatures.  The error bars shown on the 10\% Ce data
  points are more typical.
  }
  \label{fig:figure4}
\end{center}
\end{figure}

Two samples (10\% and 20\% Ce$\leftrightarrow$Gd) were measured,
and of all the doped samples studied these exhibit the most
remarkable and revealing transport properties: a large drop in
$R_H$ below $T_{Magnetic}$ (in fact becoming negative in the 10\%
sample below $\sim$30K), and a large TEP at room temperature
which, like the Hall effect, drops very rapidly below
$T_{Magnetic}$.  These data are shown in Figure \ref{fig:figure4},
along with the resistivities of the two samples.

We note first that, as with other electron doping substitutions
(Ce for Gd, La for Sr and hydrogen doping), $T_{Magnetic}$ is
driven upwards. This appears to reflect an increasing Ru$^{4+}$
fraction. The 10\% Ce sample will be dealt with first. As with the
undoped sample, the departure from cuprate-like properties below
$T_{Magnetic}$ indicates a transition from localized to itinerant
behavior in the RuO$_2$ layer.  In this case, however, the room
temperature TEP, $S_{290}=110\mu$VK$^{-1}$, indicates a much lower
CuO$_2$ layer carrier concentration of $p\sim 0.03$ holes/Cu. This
is consistent with the increased Hall coefficient, which is
probably still depressed from the true CuO$_2$ value by residual
conductivity in the RuO$_2$ layer, and the large resistivity with
its insulating upturn at low temperature.  Having concluded that
$p$, and hence $\sigma_{Cu}$, is much lower than in the undoped
sample, the reason for the dramatic effects seen in $R_H$ and $S$
below $T_{Magnetic}$ becomes clear: the ratio
$\sigma_{Ru}/\sigma_{Cu}$ is much larger in the Ce-doped sample at
low temperature, allowing the intrinsic RuO$_2$ layer properties
to dominate the behavior.

The effect of changes in $\sigma_{Ru}/\sigma_{Cu}$ is greater for
$R_H$ than $S$, but for these samples the increased $\sigma_{Ru}$
depresses the $S_{Cu}$ contribution to the total TEP so much that
$S_{Ru}$ dominates below 50K.  The basis for this assertion is the
double maximum in $dS\over dT$: initially, at low temperatures,
$S(T)$ follows a curve reasonably consistent with the TEP of
SrRuO$_3$.  This contribution would appear to saturate at a value
of $\sim 40\mu$VK$^{-1}$; however, above 50K increasing
$\sigma_{Cu}$ allows $S_{Cu}$ to contribute, and the overall TEP
then rises more rapidly.

The same qualitative treatment may be applied successfully to the
Hall effect data, though in order to explain the negative values
below $\sim$30K it is necessary to assume a negative Hall
coefficient for the RuO$_2$ layer of around $-1\times
10^{-9}$m$^3$C$^{-1}$.  Examining typical data from the
Sr$_{n+1}$Ru$_n$O$_{3n+1}$ series one finds that $R_H$ of
Sr$_3$Ru$_2$O$_7$ remains positive at all temperatures, while that
of Sr$_2$RuO$_4$ becomes negative below 20K, but reaches just
$-1\times 10^{-10}$m$^3$C$^{-1}$ near 1K. However, SrRuO$_3$,
which has the most similar ferromagnetic RuO$_2$ layer to
RuSr$_2$GdCu$_2$O$_8$, has a negative $R_H$ below 100K, reaching a
field-dependent value of $\sim -1\times 10^{-9}$m$^3$C$^{-1}$
below 60K\cite{Yang1999}. Thus the value of $R_H$ observed in the
Ce-doped sample at low temperature is the same order of magnitude
as that in SrRuO$_3$, confirming that the RuO$_2$ layer dominates
the transport properties.
It is interesting to note that, though it may not be a large
effect, Ce substitution for Gd should drive the mean Ru valence
closer to 4+, as it is in SrRuO$_3$.

\begin{figure}[htb]
\begin{center}
  \epsfig{file=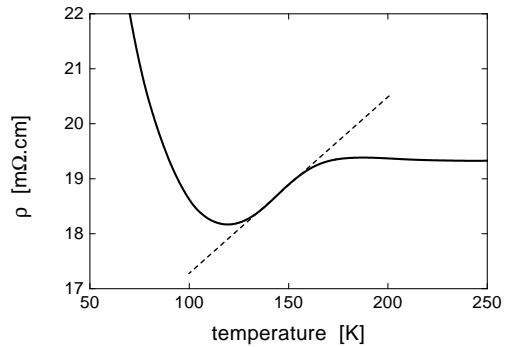, width=3.2in}
  \caption{Enlarged view of the resistivity of the 10\% Ce-doped
  RuSr$_2$GdCu$_2$O$_8$ sample showing $T$-linear resistivity below $T_{Magnetic}$.}
  \label{fig:figure5}
\end{center}
\end{figure}

Turning now to the resistivity, one encounters a problem: if the
RuO$_2$ layer is indeed metallic below $T_{Magnetic}$, why does
the resistivity increase so dramatically as $T\rightarrow 0$?
There are two possible answers to this question: either both the
RuO$_2$ and CuO$_2$ layers are at least semiconducting, but such
that $\sigma_{Ru}/\sigma_{Cu}>1$, or it may be that insulating
grain boundaries cause the upturn.  The second of these scenarios
seems more likely. In this case the TEP and Hall effect, being
much less sensitive to inter-grain connectivity, are determined by
a weakly metallic intrinsic $\sigma_{Ru}$. Support for this
conclusion is provided by close examination of the resistivity
(Figure \ref{fig:figure5}) which shows an extended metallic region
below $T_{Magnetic}$.
This type of behavior is not uncommon in RuSr$_2$GdCu$_2$O$_8$\ --
in fact extrinsic upturns in resistivity are observed in poorly
annealed undoped samples.  Interestingly though, transport
measurements on SrRuO$_3$ also show a minimum in resistivity below
T$_{\mathrm{Curie}}$ in samples where there is some disorder in
the RuO$_2$ layer\cite{Klein1996}. The temperature at which the
minimum occurs, and the magnitude of the upturn below it both
increase with RuO$_2$ layer disorder: in good quality films the
highest temperature minimum observed is 40K, coincident with the
maximum residual resistivity\cite{Klein1996}. As the RuO$_2$ layer
may be considerably disordered in these doped
RuSr$_2$GdCu$_2$O$_8$ samples, it is possible that the behavior
shown in Figure \ref{fig:figure5} is intrinsic to the RuO$_2$
layer, and not a product of grain boundary resistivity. An
extended anneal of the Ce-doped sample could be used to test the
origin of the effect, as it should improve the quality of the
grain boundaries in the material.

The 20\% Ce-doped sample appears initially to be inconsistent with
the interpretation of the 10\% Ce data presented above: $R_H$ is
higher at room temperature, as one would expect for even greater
under-doping caused by the increase in Ce content, but $S_{290}$
is actually \textit{lower} than that of the 10\% sample,
apparently implying an increased hole concentration. The paradox
may be resolved straightforwardly if one accepts that
$\sigma_{Ru}\approx\sigma_{Cu}$, even at room temperature, in this
sample.  Then the values of $S_{290}$ and $R_H(290)$ for the
material are no longer dominated by the CuO$_2$ layer;
$S_{290}^{Cu}$ and $R_H^{Cu}(290)$ may both in fact be much larger
than the overall measured values so long as $S_{290}^{Ru}$ and
$R_H^{Ru}(290)$ are lower, as expected.  From the peak in the Hall
effect data, $T_{Magnetic}$ of this sample appears to be depressed
by about 30K from its value in the undoped sample, while the
resistivity at 50K is more than two orders of magnitude greater,
and the sample is insulating, probably due to extreme granularity.

\subsubsection*{Calculation of $\beta(T)$ in Ce-doped RuSr$_2$GdCu$_2$O$_8$}
The ratio $\beta(T)={\sigma_{Ru}(T)/\sigma_{Cu}(T)}$ may be
extracted from the data for the 10\% Ce doped sample using the
two-band model, as described for the undoped material in the
previous Chapter. As in the previous calculation, typical
$S^{Cu}(T)$ and $R_H^{Cu}(T)$ data are matched to the high
temperature RuSr$_2$GdCu$_2$O$_8$ data, where $\sigma_{Ru}$ is
assumed to be small compared with $\sigma_{Cu}$ and the overall
properties reflect those of the CuO$_2$ layer most strongly.  The
deviation from cuprate-like behavior at lower temperatures is then
used to extract the ratio $\beta(T)$.

For this sample, $S^{Cu}(T)$ data were taken as 1.05 times $S(T)$
measured on a sample of under-doped sintered
YBa$_2$Cu$_3$O$_{7-\delta}$, with $\delta \simeq
0.6$\cite{Cooper1996}. $R_H^{Cu}(T)$ data were taken as 1.16 times
$R_H$ measured on a similar sample with $\delta \simeq
0.62$\cite{Carringtonthesis}. $R_H$ and $S$ are particularly
strong functions of doping in this region of the phase diagram:
the good agreement in the values of $\delta$ required for the two
sets of data to match those of RuSr$_2$GdCu$_2$O$_8$ suggests that
the assumption of negligible $\sigma_{Ru}$ is reasonable.
\begin{figure}[htb]
\begin{center}
  \epsfig{file=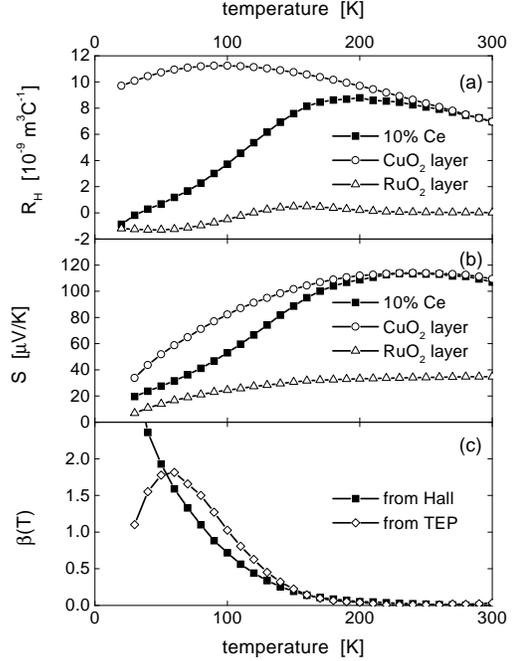, width=3.2in} \caption{(a) Hall effect
  and (b) thermopower data for
  RuSr$_2$Gd$_{0.9}$Ce$_{0.1}$Cu$_2$O$_8$, together with the
  estimated RuO$_2$ and CuO$_2$ layer values, as described in the
  text. Panel (c) shows the ratio
  $\beta={\sigma_{Ru}\over\sigma_{Cu}}$ calculated in the two-band
  model from thermopower and, independently, from Hall effect.}
  \label{fig:figure6}
\end{center}
\end{figure}
The TEP of the RuO$_2$ layer is approximated by that of sintered
SrRuO$_3$, as before. As the Hall effect becomes negative at low
temperatures in Ce-doped RuSr$_2$GdCu$_2$O$_8$ taking
$R_H^{Ru}\simeq 0$, as was done for the undoped material, will not
work.  Instead a rough approximation to data for SrRuO$_3$ is
used\cite{Yang1999}, which shows a field-dependent value of $\sim
-1\times 10^{-9}$ at 20K.

The measured and estimated data, together with the results of the
calculations are shown in Figure \ref{fig:figure6}.  Above 50K
there is remarkable agreement between $\beta(T)$ calculated from
the TEP data ($\beta_{TEP}$) and that calculated independently
from the Hall effect data ($\beta_{Hall}$), lending confidence
both to the model and to the estimated $R_H(T)$ and $S(T)$ data
for the RuO$_2$ and CuO$_2$ layers. Below 50K the agreement is not
so good: $\beta_{Hall}$ carries on increasing, a direct result of
$R_H^{1212}$ becoming very close to the estimated $R_H^{Ru}$ at
low temperatures. $S^{1212}$ does not approach the estimated
$S^{Ru}$ as closely, and hence $\beta_{TEP}$ does not continue to
increase. Emerging clearly from these data is a large increase in
$\beta$ below $T_{Magnetic}$.  At 50K $\sigma_{Ru}/\sigma_{Cu}\sim
1.9$, whereas for the undoped material the increase in
$\sigma_{Ru}/\sigma_{Cu}$ is just 0.3. The properties of the
RuO$_2$ layer dominate the overall transport properties of
RuSr$_2$GdCu$_2$O$_8$ below $\sim 90$K in this 10\% Ce-doped
sample.

\subsubsection*{Other Doped Samples} \label{sect:otherdoped} 
\begin{figure}[htb]
\begin{center}
  \epsfig{file=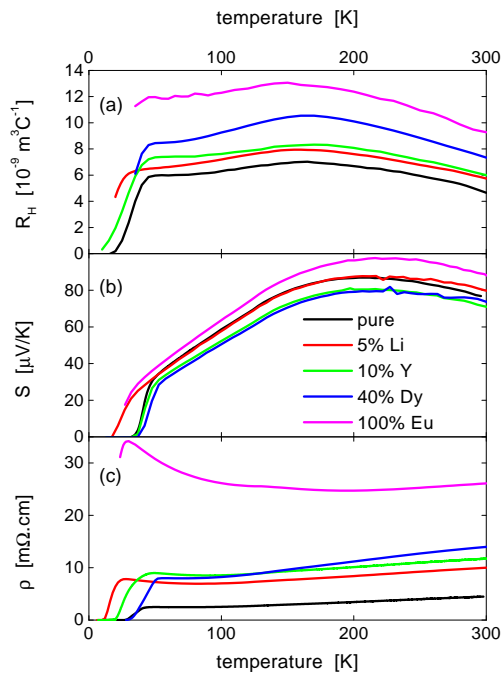, width=3.2in}
  \caption{(a) Hall effect, $R_H(T)$, (b) thermopower, $S(T)$, and
  (c) resistivity, $\rho(T)$, data for RuSr$_2$GdCu$_2$O$_8$ with
  10\%Y, 40\%Dy and 100\%Eu substituted for Gd, or 5\%Li
  substituted for Cu.} \label{fig:figure7}
\end{center}
\end{figure}

The
remainder of the doped samples studied contained Y, Dy and Eu on
the Gd site, plus a 5\% Li-doped sample, in which Cu is
substituted. The transport data for all these samples are shown in
Figure \ref{fig:figure7}.
The Hall effect shows the `usual' anomalous downturn below
$T_{Magnetic}$ in all these samples.  The magnitude of the
downturn, due to the transition to a more itinerant Ru layer, is
approximately constant, leading to the conclusion that doping the
Cu and Gd sites does not greatly affect the localization of
carriers in the RuO$_2$ layer.

Substituting a small amount of Li$^+$ for Cu$^{2+}$ causes
virtually no change in the TEP, but depresses $T_c$ by
approximately 20K. The Hall effect of this sample is slightly
larger than that of the undoped sample, possibly due to some
cross-substitution of Li with Ru, depressing $\sigma_{Ru}$, or a
slight decrease in the CuO$_2$ layer carrier concentration. These
results are consistent with Li$^+$ acting as a pair-breaker in the
CuO$_2$ layer, but with little other effect on transport
properties. The rate of suppression of $T_c$ with Li substitution
in RuSr$_2$GdCu$_2$O$_8$, $\sim 4$K/\%, is about one quarter of
that observed in under-doped YBa$_2$Cu$_3$O$_{7-\delta}$ ($\delta
= 0.4$) when either Li or Zn is substituted for Cu
\cite{Bobroff1999}.  However the concentration of Li in the
RuSr$_2$GdCu$_2$O$_8$ sample studied is nominal, and the
difference in the rate of suppression may simply reflect loss of
Li by vaporization during the long synthesis and anneal.

The isovalent substitution of Y or Dy for Gd actually causes a
slight \textit{decrease} in the TEP of RuSr$_2$GdCu$_2$O$_8$,
these being the only substitutions studied to do so.  The implied
increase in the doping level of the CuO$_2$ layers, presumed to
arise from an ion-size effect, is confirmed by the increased $T_c$
in these samples - approximately 2K (10\% Y) and 6K (40\% Dy)
higher than in the undoped sample as seen by both resistivity and
TEP measurements. However, the magnitude of the Hall effect is
larger for these samples than for the undoped ones. Having argued
that the CuO$_2$ layer is less under-doped in these two samples,
this effect may only arise from a decrease in the conductivity of
the RuO$_2$ layer, partially removing the `shorting' of the
CuO$_2$ layer Hall effect.

Full substitution of Eu for Gd causes an increase in $S_{290}$ to
90$\mu$VK$^{-1}$, $T_c$ as measured by the TEP or resistivity
drops significantly, and $R_H$ is greatly increased. All these
results suggest a drop in the CuO$_2$ layer hole concentration,
again consistent with the above-noted ion-size effect, perhaps
coupled with a decrease in $\sigma_{Ru}$. The resistivity of this
sample is much less metallic than that of the others (which are
all metallic, with magnitudes two to three times the well-annealed
undoped sample), consistent with this interpretation.

\section*{Discussion}
The Hall effect and TEP data described in this paper provide
strong evidence for a transition from very poorly metallic to more
itinerant behavior in the RuO$_2$ layer below $T_{Magnetic}$.
Results from substituted RuSr$_2$GdCu$_2$O$_8$ samples confirm
this picture. The universal relationship between $S_{290}$ and
$p_{Cu}$ appears to hold in RuSr$_2$GdCu$_2$O$_8$ as a result of
the low $\sigma_{Ru}$ at room temperature, though below
$T_{Magnetic}$ both $S$ and $R_H$ are reduced by `shorting' of
$S_{Cu}$ by the RuO$_2$ layer.

The two-band model proposed is successful in explaining most of
the existing data qualitatively: the anomalies, which lie mainly
in resistivity data, are most likely due to grain boundary and
density effects. The quantitative agreement is also reasonably
good.  The results support a picture in which the RuO$_2$ layer in
the pure compound is localized above $T_{Magnetic}$, with
$\sigma_{Ru}\sim 0.1\sigma_{Cu}$, but becomes more conducting
below $T_{Magnetic}$, mirroring the behavior of other ruthenates.

The transition from localized to itinerant RuO$_2$ layer behavior
at $T_{Magnetic}$ in the undoped compound may be modified by
substituting Ru with Sn or Nb.  Sn increases the doping level of
the CuO$_2$ layers, raising $T_c$ and suppressing $T_{Magnetic}$,
and simultaneously drives the RuO$_2$ layer more insulating.  Nb
under-dopes the CuO$_2$ layers, lowering $T_c$, and also appears
to drive the RuO$_2$ layer insulating, though the 20\% sample does
not show the expected semiconducting resistivity. These results
imply an initial Ru valence lying
between 4+ and 5+, in agreement with 
X-ray Absorption Near Edge Spectroscopy (XANES) data which may be
modelled as an admixture of Ru$^{4+}$ and Ru$^{5+}$ spectra in a
40:60 ratio\cite{Liu2000}.

As might be expected, doping of the Cu site has little effect on
$T_{Magnetic}$ or the transport properties of the RuO$_2$ layer.
Li$^+$ acts as a pair-breaking impurity in the CuO$_2$ layer and
causes a depression of $T_c$ in line with its behavior in other
cuprates. Isovalent doping of the Gd site with other lanthanide
elements changes the CuO$_2$ layer doping level, with a remarkably
strong variation in $T_c$.  This appears to be a ion-size doping
effect. Altervalent substitution of Ce for Gd rapidly reduces the
doping level of the CuO$_2$ layers and drives the material
non-superconducting.  In all but the Ce-doped samples, the
conductivity of the RuO$_2$ layer only ever reaches a modest
fraction of that of the CuO$_2$ layer.  In the 10\% Ce-doped
sample the more heavily under-doped CuO$_2$ layer has an
insulating upturn at low temperature, whilst the RuO$_2$ layer
remains more metallic, and so the ratio of their conductivities
reaches at least 1.9.

\section*{Conclusions}
To a first approximation the electronic properties of the CuO$_2$
layer in RuSr$_2$GdCu$_2$O$_8$ are the same as those of similar
CuO$_2$ layers in other high-$T_c$ cuprate superconductors in all
respects.  This conclusion is supported by the resistivity, TEP
and Hall effect data presented here, and by results on the
specific heat jump at $T_c$. On a more detailed level,
magneto-transport measurements reveal an interaction between the
carriers in these layers and the magnetization of the RuO$_2$
layer. This interaction, with an energy which would seem to be of
the same order as the SC energy gap\cite{McCrone1999}, is not
sufficient to destroy superconductivity.

The electronic properties of the RuO$_2$ layer appear to bear a
remarkable similarity to those observed in the ruthenate
SrRuO$_3$.  At room temperature the conductivity of the layer is
perhaps 10\% of that of the CuO$_2$ layer, with $k_Fl_{Ru}\approx
0.2$, indicating very badly metallic or localized behavior. Below
$T_{Magnetic}$ the conductivity of the layer rises significantly -
by at least 0.3$\sigma_{Cu}$.  This increase raises the weighting
of the RuO$_2$ layer properties relative to those of the CuO$_2$
layer in the admixture that determines the overall transport
properties of RuSr$_2$GdCu$_2$O$_8$.  As the Hall effect and TEP
of the RuO$_2$ layer are both considerably smaller than those in
the CuO$_2$ layer the result is a drop in both $R_H$ and $S$ below
$T_{Magnetic}$.  In pure RuSr$_2$GdCu$_2$O$_8$, and also in most
of the substituted variants studied, $\sigma_{Ru}$ remains lower
than $\sigma_{Cu}$ over the whole temperature range. For the
Ce-doped samples studied, however, the CuO$_2$ layer becomes
insulating at low temperatures, allowing the poorly metallic
RuO$_2$ layer to dominate the conductivity, and its intrinsic
transport properties to show strongly in the overall $R_H$ and $S$
of the material.

The two-band model of parallel conduction in the RuO$_2$ and
CuO$_2$ layers has been very successful in modelling the transport
properties observed in all the RuSr$_2$GdCu$_2$O$_8$ samples
studied, and it has been possible to describe well the effects of
doping the different atomic sites. The inferred mixed valency of
Ru, together with the onset of itinerancy at the magnetic
transition suggests a possible role of a double-exchange mechanism
in the magnetic interactions but also raises the possibility of
charge ordering in these compounds at appropriate doping levels.

This work was supported by the U.K. Engineering and Physical
Sciences Research Council and the New Zealand Marsden Fund for
research (JLT) and travel funds (CB).

\end{document}